\documentclass[a4paper]{particles2019}

\usepackage[utf8]{inputenc}
\usepackage[english]{babel}
\usepackage{xcolor}
\usepackage{graphicx}
\usepackage{amsmath}
\usepackage{amsfonts}
\usepackage{amssymb}

\title{MONODISPERSE GAS-SOLID MIXTURES WITH INTENSE INTERPHASE INTERACTION IN TWO-FLUID SMOOTHED PARTICLE HYDRODYNAMICS}

\author{OLGA P. STOYANOVSKAYA$^{1,2}$, TATIANA A. GLUSHKO$^{2}$, \\
VALERY N. SNYTNIKOV$^{2,3}$ AND NICOLAY V. SNYTNIKOV$^{4}$}

\heading{Olga P. Stoyanovskaya, Tatiana A. Glushko, Valeriy N. Snytnikov and Nikolay V. Snytnikov}

\address{$^{1}$ Lavrentiev Institute of Hydrodynamics SB RAS\\
630090, Novosibirsk, Russia\\
Web page: http://www.ict.nsc.ru/en
\and
$^{2}$ Novosibirks State University\\
630090, Novosibirsk, Russia\\ Web page: https://english.nsu.ru
\and
$^{3}$ Boreskov Institute of Catalysis SB RAS\\
630090, Novosibirsk, Russia\\

Web page: http://en.catalysis.ru/
\and
$^{4}$ Institute of Computational Mathematics and Mathematical Geophysics\\
630090, Novosibirsk, Russia\\
e-mail: nik@ssd.sscc.ru - Web page: https://icmmg.nsc.ru/en}


\abstract{Simulations of gas-solid mixtures are used in many scientific  and industrial applications. Two-Fluid Smoothed Particle Hydrodynamics (TFSPH) is an approach when gas and solids are simulated with different sets of particles interacting via drag force. Several methods are developed for computing drag force between gas and solid grains for TFSPH. 

Computationally challenging are simulations of gas-dust mixtures with intense intephase interaction, when velocity relaxation time $t_{\rm stop}$ is much smaller than dynamical time of the problem. In explicit schemes the time step $\tau$ must be less than $t_{\rm stop}$, that leads to high computational costs. Moreover, it is known that for stiff problems both grid-based and particle methods may require unaffordably detailed resolution to capture the asymptotical bahaiviour of the solution. To address this problem we developed fast and robust method for computing stiff and mild drag force in gas solid-mixtures based on the ideas of Particle-in-Cell approach. In the paper we compare the results of new and previously developed methods on test problems.}

\begin{document}

\section{INTRODUCTION}

In the paper two-fluid model of gas-solid mixture is discussed. In this model gas is considered as a carrier phase and dust grains are considered as dispersed phase. It is assumed that the solid phase has one typical size in each volume (so this phase is monodisperse). Therefore, the continuity and motion equations for gas and dispersed phase have the following form:


\begin{equation}
\label{eq:gas}
\displaystyle\frac{\partial \rho_{\rm g}}{\partial t}+\nabla (\rho_{\rm g} v)=S_{\rm g},\ \ \ 
 \rho_{\rm g} \left[\displaystyle\frac{\partial v}{\partial t}+(v \cdot \nabla) v \right]=-\nabla p+ \rho_{\rm g} g - f_{\rm drag}+f_{\rm g},
\end{equation}
\begin{equation}
\label{eq:dust}
\displaystyle\frac{\partial \rho_{\rm d}}{\partial t}+\nabla (\rho_{\rm d} u)=S_{\rm d},\ \ \ 
\rho_{\rm d} \left[\displaystyle\frac{\partial u}{\partial t}+(u \cdot \nabla) u \right]=\rho_{\rm d} g + f_{\rm drag}+f_{\rm d},
\end{equation}
where $\rho_{\rm g}$ and $\rho_{\rm d}$ are volume density of gas and dust, $v$ and $u$ are velocities of gas and dust, $p$ is pressure, $g$ is gravity acceleration, $S_{\rm g},S_{\rm d}$ are sourses and sinks for gas and dust, $f_{\rm g},f_{\rm d}$ are forces affecting gas and dust except for pressure, gravitaion and drag, $f_{\rm drag}$ is drag force per unit volume:
\begin{equation}
\label{eq:fdrag}
f_{\rm drag}=\rho_{\rm d} \displaystyle \frac{v-u}{t_{\rm stop}},
\end{equation}
where $t_{\rm stop}=t_{\rm stop}(a,\rho_{\rm g}, c_s, v-u)$ is a velocity relaxation time. Here $a$ is particle size, $c_s$ is the sound speed in gas. In this work we consider particular case  when $t_{\rm stop}$ does not depend on $v-u$ that corresponds to Epstein and Stokes regimes (see details, e.g. in \cite{Saito2003}).


Computing of pure gas dymanics (solution of (\ref{eq:gas}) with $f_{\rm drag}=0$) using explicit schemes requires time step $\tau$ that satisfies Courant condition:
\begin{equation}
\label{eq:Courant}
 \tau < \displaystyle \mathrm{CFL} \frac{h}{\max (v,c_s)},
\end{equation}
where $\mathrm{CFL}<1$ is the Courant limeter. Additional necessary condition arises during computing of dusty gas dynamisc using explicit schemes
\begin{equation}
\label{eq:tauStiff}
 \tau < t_{\rm stop}.
\end{equation}
Violation of this condition leads to numerical instability. Condition (\ref{eq:tauStiff}) is extremally prohibitive for intense interphase interaction (for small $t_{\rm stop}$). Intense interphase interaction arises in many applications of gas-particle mixtures (e.g. in modelling reactors with finely-dispersed catalyst, in planet formation from gas-dust circumstellar disks etc.) and is characterized by the fact that time of momentum transfer is much less than process time.


In the case of intense interphase interaction, $f_{\rm drag}$ is called stiff relaxation term \cite{JinLivermore}. Effective schemes for problems with stiff relaxation terms are designed in way that they could preserve asymptotical solution even with $\tau \gg t_{\rm stop}$. Main ideas of such design are described in \cite{Jin2012,Albi2019} and assosiated with using of implicit approximation of stiff relaxation term along with explisit approximation of other terms. This approaches are developed for euler methods for solving fluid dynamics equations and are used in different applications e.g. \cite{StoyanovskayaDust,Multiscale}. Transfering this ideas to lagrangian methods (for example, to TFSPH in which gas and dust are modelled by different sets of particles) encounter new difficulties caused by the fact that carrier and disperse phases quantities are known in different points of space. For TFSPH authors of \cite{LaibePrice2011,BateDust2014,SPHIDIC} shown that the way of interpolation this parametres for computing $f_{\rm drag}$ influences the method's property to preserve asymptotic. In particular, \cite{LaibePrice2011} demonstrated that classical method for computing drag force in TFSPH \cite{MonaghanKocharyan1995} captures the asymptotic of the solution for small $t_{\rm stop}$ only with
\begin{equation}
\label{eq:hLaibePrice}
h< c_s t_{\rm stop}.
\end{equation}

Getting over this restriction is crucial for modelling mixtures with intense interphase interaction. For this reason authors of \cite{BateDust2014}, \cite{SPHIDIC} proposed other methods to computing drag forces. In this paper we present quantitative comparison of the classical \cite{MonaghanKocharyan1995} and new \cite{BateDust2014}, \cite{SPHIDIC} approaches focusing on their ability to preserve asymptotic properties of the solution. The methods are described in detail in section \ref{sec:sph}, the test problem results are given in section \ref{sec:results} and the summary is provided in section \ref{sec:summary}. 

\section{METHODS FOR COMPUTING DRAG TERMS IN TWO-FLUID\\
SMOOTHED PARTICLE HYDRODYNAMICS}
\label{sec:sph}

Let us rewrite the motion equations in (\ref{eq:gas})-(\ref{eq:dust}) assuming $K=\displaystyle\frac{\rho_{\rm d}}{t_{\rm stop}}$:
\begin{equation}
\label{eq:systemK}
\left\{
 \begin{array}{lcl}
        \displaystyle 
        \frac{\mathrm{d}v}{\mathrm{dt}} = -\displaystyle \frac{\nabla p}{\rho_{\rm g}} + g -\displaystyle\frac{K}{\rho_{\rm g}} (v - u), \\
        \displaystyle 
        \frac{\mathrm{d}u}{\mathrm{dt}} = g + \displaystyle\frac{K}{\rho_{\rm d}} (v - u). 
    \end{array}
\right.
\end{equation}

Further we will give the schemes for solving the equations of gas and dust motion (\ref{eq:systemK}) in the standard SPH notation. We will consider only the schemes in  which gas and dust are simulated by different sets of particles, i.e. by the two-fluid approach for smoothed particle hydrodynamics (TFSPH). Let $n$ be the number of the time step. Following the notation introduced in \cite{MonaghanKocharyan1995}, we will use $a,b$ as the indices for gas particles, and $j,k$ as the indices for dust particles.

\subsection{The MK Monaghan–Kocharyan explicit scheme.}
A method for computing the drag force, which was proposed in \cite{MonaghanKocharyan1995} (hereinafter referred to as MK (Monaghan-Kocharyan Drag)), is classical for smoothed particle hydrodynamics. This method is based on the computing of the relative velocity between each pair of gas-dust particles and is employed in astrophysical and engineering applications of two-phase medium mechanics \cite{Maddison2004TFSPH,FranceDustCode,Gonzalez2017,TFSPH} and others.

We implemented this scheme so that the summand accounting for drag uses the velocities from the previous time step: 
\begin{equation}
\displaystyle\frac{\mathrm{d}v^n_a}{\mathrm{dt}}= - m_{\rm g} \sum_b \left(\frac{p_b}{(\rho^n_{b,\rm{g}})^2} + \frac{p_a}{(\rho^n_{a,\rm g})^2} \right) \bigtriangledown_a W^{n}_{ab} 
- \sigma m_{\rm d} \sum_j \frac{K_{aj}}{\rho^n_{a, \rm g} \rho^n_{j,\rm d}} \frac{(v_a^{n} - u_j^{n},r_{ja})}{r_{ja}^2+\eta^2}r_{ja}W^{n}_{ja}
+g_a,
\label{eq:MonKoch1995v}
\end{equation}
\begin{equation}
\displaystyle\frac{\mathrm{d}u^n_j}{\mathrm{dt}}= \sigma m_{\rm g} \sum_a \frac{K_{aj}}{\rho^n_{a, \rm g} \rho^n_{j,\rm d}} \frac{(v_a^{n} - u_j^{n},r_{ja})}{r_{ja}^2+\eta^2}r_{ja}W^{n}_{ja}+g_j,
\label{eq:MonKoch1995u}
\end{equation}
\begin{equation}
K_{aj}=\displaystyle\frac{\rho^n_{j,\rm d} \rho^n_{a, \rm g} c^n_{a,\rm s}}{s_j^n \rho^n_{j,\rm s}},
\end{equation}
where $m_{\rm g}$ and $m_{\rm d}$ are the masses of gas and dust particles, respectively, $r_{ja} = r_j - r_a$, $\eta$ is a clipping constant, $\eta^2 = 0.001 h^2$, $s_j$ is the radius of a spherical dust particle with index $j$, $\sigma$ is the constant determined by dimensionality of the problem (for one-dimensional problems, $\sigma=1$), and $W^{n}_{ab}=W(h,r_{ab})$ is the smoothing kernel.

The MK scheme (\ref{eq:MonKoch1995v})-(\ref{eq:MonKoch1995u}) of the first order approximation with respect to time satisfies the momentum conservation law in the entire computational domain, which means that the momentum lost by gas due to drag on dust completely coincides with the momentum acquired by dust due to drag on gas.

\subsection{The semi-implicit ISPH scheme with interpolation of the first order approximation with respect to time.}

The second method for computing the drag consists in the calculation of gas characteristics at the points where dust particles are located (and vice versa) using the SPH interpolation formulas:
\begin{equation}
\label{eq:VUvolaverage}
v^n_j=\displaystyle m_{\rm g} \sum_a \frac{v^n_a}{\rho^n_{a, \rm g}} W^n_{aj}, \quad u^n_a = \displaystyle m_{\rm d} \sum_j \frac{u^n_j}{\rho^n_{j, \rm d}} W^n_{ja}, 
\end{equation}
where $u_a$ is the dust velocity at a spatial point where the gas particle with $a$ index  is located, and $v_j$ is the gas velocity at a spatial point where the dust particle with $j$ index is located.

As a result, all features of the gas-dust medium become known for each model particle. This method and its modifications are applied in refs. \cite{BateDust2014,ClarkeDust2015,RiceEtAl2004}. We apply this idea to construct a semi-implicit scheme that would not require the fulfillment of condition (\ref{eq:tauStiff}) for obtaining stable solutions. In particular, parsimonious is the following ISPH scheme with the first order approximation with respect to time (the quantities calculated using interpolation formulas are marked in blue, while the quantities derived from those calculated by interpolation formulas are marked in red):
\begin{equation}
\label{eq:VISPH}
\displaystyle\frac{v^{n+1}_a-v^n_a}{\tau}= - \sum_b m_b\left(\frac{p_b}{(\rho^n_{b,\rm{g}})^2} + \frac{p_a}{(\rho^n_{a,\rm g})^2} \right) \bigtriangledown_a W^{n}_{ab} - 
\frac{\textcolor{red}{K_a^n}}{\rho^n_{a,\rm g}} (v_a^{n+1} - {u_{a}^{n+1}})+g_a,
\end{equation}
\begin{equation}
\displaystyle\frac{u^{n+1}_a-\textcolor{blue}{u^n_a}}{\tau}= \frac{\textcolor{red}{K_a^n}}{\textcolor{blue}{\rho^n_{a,\rm d}}} (v^{n+1}_a - u^{n+1}_a)+g_a.
\end{equation}
\begin{equation}
\label{eq:UISPH}
\displaystyle\frac{v^{n+1}_j-\textcolor{blue}{v^n_j}}{\tau}= - \sum_i m_i\left(\textcolor{blue}{\frac{p_i}{(\rho^n_{i,\rm{g}})^2} + \frac{p_j}{(\rho^n_{j,\rm g})^2}} \right) \bigtriangledown_j W^{n}_{ij} - 
\frac{\textcolor{red}{K_j^n}}{\textcolor{blue}{\rho^n_{j,\rm g}}} (v_j^{n+1} - {u_{j}^{n+1}})+g_j,
\end{equation}
\begin{equation}
\displaystyle\frac{u^{n+1}_j-u^n_j}{\tau}= \frac{\textcolor{red}{K_j^n}}{\rho^n_{a,\rm d}} ({v^{n+1}_j} - u^{n+1}_j)+g_j.
\end{equation}
\begin{equation}
\label{eq:KintSPH}
\textcolor{red}{K^n_a}=\displaystyle\frac{\rho^n_{a, \rm g} c^n_{a,\rm s}}{\textcolor{blue}{s_a^n \rho^n_{a,\rm s}}}, \quad 
\textcolor{red}{K^n_j}=\displaystyle\frac{\textcolor{blue}{\rho^n_{j, \rm g} c^n_{j,\rm s}}}{s_j^n \rho^n_{j,\rm s}}.
\end{equation}


\subsection{A new SPH-IDIC scheme -- the implicit `drag in cell''.}

In addition, computing of the drag force can be based on the idea of the particle-in-cell method for simulation of gas-dust flows \cite{Andrews1996}. The parsimonious semi-implicit SPH-IDIC approach based on this idea was suggested and tested in our earlier paper \cite{SPHIDIC}. A detailed description of this approach is presented below. 

At each time instant, we will decompose the entire calculation region into disjoint volumes so that the merging of these volumes will coincide with the entire region. Suppose a separate volume contains $N$ gas particles of a similar mass $m_{\rm g}$ and $L$ dust particles of a similar mass $m_{\rm d}$, with $N>0$, $L>0$. Introduce the volume-averaged values of $t^*_{\rm stop}$ and $\rho^*_{\rm d}$ (anywise) and assume that 
\begin{equation}
\label{eq:aveEpsilon}
\varepsilon^*=\displaystyle\frac{m_{\rm d} L}{m_{\rm g} N}, 
\end{equation}
thus determining 
\begin{equation}
K^*=\frac{\rho^*_{\rm d}}{t^*_{\rm stop}}, \ \ \rho^*_{\rm g}=\frac{\rho^*_{\rm d}}{\varepsilon^*}. 
\end{equation}

Let us assume that in computing the drag force that acts from gas on dust, the gas velocity is constant over the entire volume and equal to $v_*$, whereas dust particles have different velocities (and vice versa). 
In addition, we will calculate the drag factor and density using values of the quantities from the preceding time layer, and relative velocity -- from the subsequent layer. The resulting scheme will have the form 
\begin{equation}
\displaystyle\frac{\mathrm{d}v^n_a}{\mathrm{dt}}= - \sum_b m_b\left(\frac{p_b}{(\rho^n_{b,\rm{g}})^2} + \frac{p_a}{(\rho^n_{a,\rm g})^2} \right) \bigtriangledown_a W^{n}_{ab} - \frac{K^*}{\rho^{n,*}_{\rm g}} (v_a^{n+1} - u_{*}^{n+1})+g_a,
\label{eq:DragInCellv}
\end{equation}
\begin{equation}
\displaystyle\frac{\mathrm{d}u^n_j}{\mathrm{dt}}= \frac{K^*}{\rho^{n,*}_{\rm d}} (v^{n+1}_* - u^{n+1}_j)+g_j,
\label{eq:DragInCellu}
\end{equation}
\begin{equation}
\label{eq:VUvolaverage}
v_*=\displaystyle\frac{\sum_{a=1}^N v_a}{N}, \quad u_*=\displaystyle\frac{\sum_{j=1}^L u_j}{L}.
\end{equation}

If the time derivative in (\ref{eq:DragInCellv})-(\ref{eq:DragInCellu}) is approximated to the first order, there exists a parsimonious method to calculate $u^{n+1}$ $v^{n+1}$. In \cite{SPHIDIC}, it was shown that the semi-implicit scheme (\ref{eq:DragInCellv})-(\ref{eq:VUvolaverage}) with the first order approximation with respect to time satisfies the momentum conservation law for each cell, i.e. the momentum lost by gas due to drag on dust completely coincides with the momentum acquired by dust due to drag on gas.

\section{RESULTS}

To measure the ability of the schemes to preserve asymptotical properties of the solution in case of intense interphase interaction two well-known test problems with available reference solution are used. The first is a problem of sound-wave propagation in isothermal gas-dust mixture. This problem has a smooth solution and suits perfectly to study the way of drag computing in SPH. We will refer to this problem as Dustywave. The second is a problem of shock wave propagation in coupled gas-dust mixture with initial conditions known as Sod shock tube. Due to discontinuity of the solution and complexity of wave structure this problem is common and challenging test for computational gas dynamics. For this problem Dustyshock name is reserved. Both Dustywave and Dustyshock problems are described in detail \cite{SPHIDIC}. Moreover, for all numerical experiments in the paper we take the same physical and numerical parameters as in \cite{SPHIDIC}. In particular, for both problems we take high drag coefficient and high concentration of dust in gas  
\begin{equation}
\label{eq:params}
   K=500, \quad \quad \displaystyle\frac{\rho_{\rm d}}{\rho_{\rm g}}=1, 
\end{equation}
which is guarantee intense interphase interaction and challenge for simulation.

\begin{figure}
    \centering
    \includegraphics[scale=0.17]{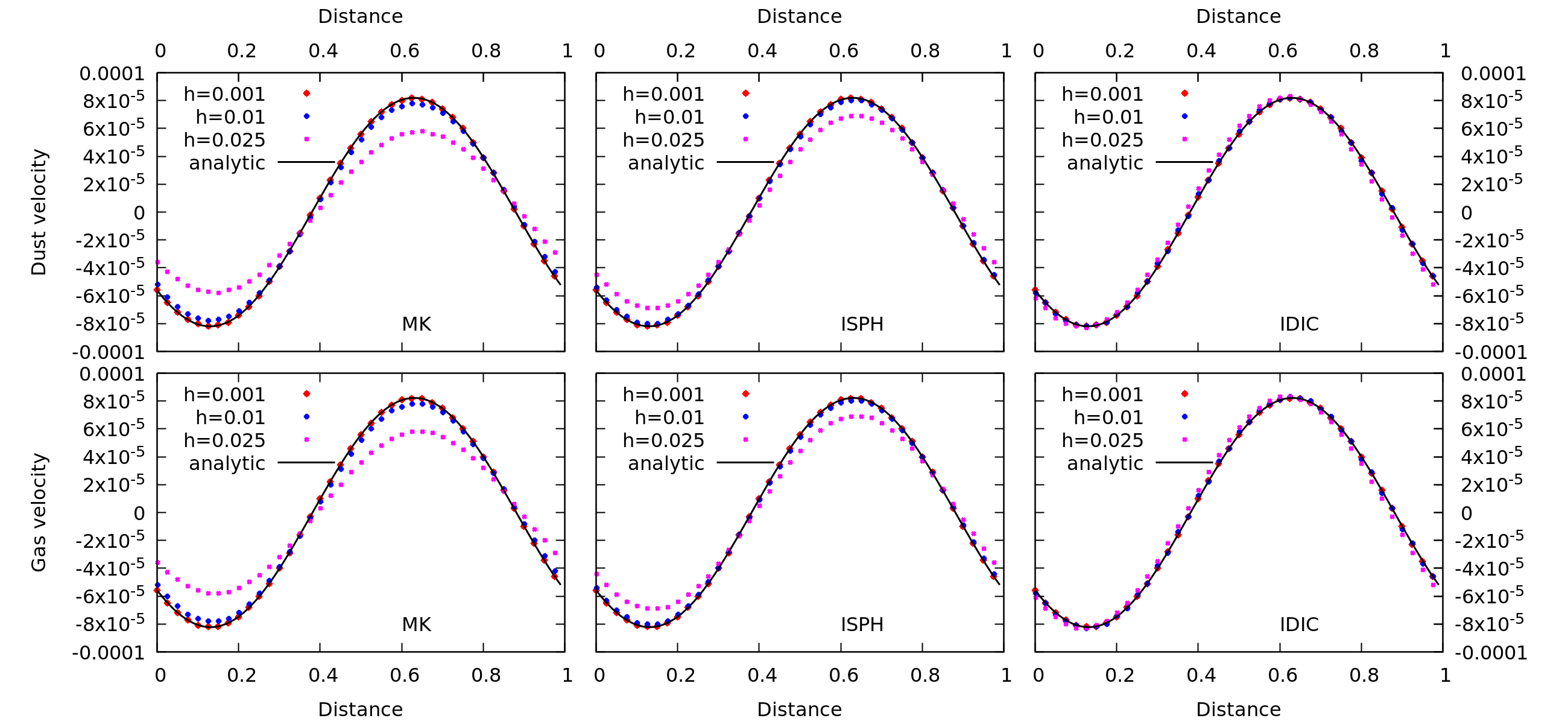}
    \caption{Solution of the DustyWave problem at the time instant $t=0.5$ found with MK (left panels), ISPH (central panels) and IDIC (right panels) methods. Relaxation time of the dust velocity with respect to gas is $t_{\rm stop}=0.002$, i.e. $t_{\rm stop} c_{\rm s} / l \ll 1$, where $l$ is the length of the computational domain. Solid black line corresponds to the analytical solution, and individual dots are the numerical solutions. At $h = 0.025$ and $h=0.01$, the time step $\tau = 0.001<t_{\rm stop}$ and the number of particles $N_{total} = 2 \times 600$ are used; at $h=0.001$, the step is $\tau = 0.0001$ and $N_{total} = 2 \times 6000$.}
    \label{fig:Dustywave}
\end{figure}

Fig.~\ref{fig:Dustywave} shows the solution of Dustywave problem at the time moment $t = 0.5$ with MK, ISPH, IDIC methods. In this case (\ref{eq:params}) leads to $t_{\rm stop} = 0.002$. The figure displays the dust and gas velocity obtained with different smoothing lengths $h$. The left panels show the results of computing for explicit schemes MK. At $h = 0.025, h=0.01$, the time step $\tau = 0.001<t_{\rm stop}$ and the number of particles $N_{total} = 2 \times 600$ are used; while at $h=0.001$, the step $\tau = 0.0001$ and $N_{total} = 2 \times 6000$. The middle and right panels present the results of computing for semi-implicit schemes ISPH and IDIC with $\mathrm{CFL}=0.1$; in this case, the number of particles was the same as for the left panels. 

\begin{figure}
    \centering
    \includegraphics[scale=0.17]{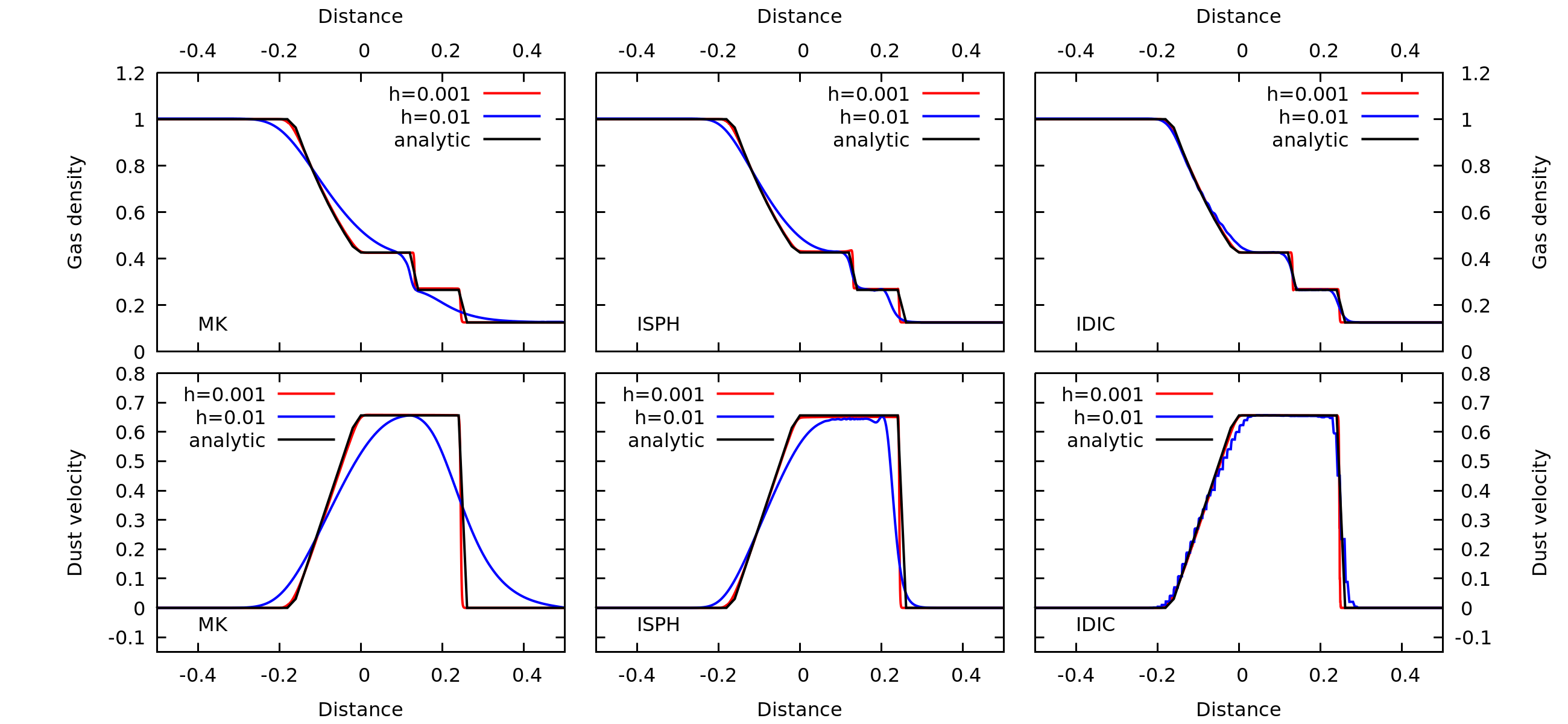}
    \caption{Solution of the DustyShock problem at the time instant $t=0.2$ found with MK (left panels), ISPH (central panels) and IDIC (right panels) methods. Relaxation time of the dust velocity with respect to gas $t_{\rm stop}=0.00025$, i.e. $t_{\rm stop} c_{\rm s} / l \ll 1$, where $l$ is the length of the computational domain. Solid black line corresponds to the analytical solution, color lines are the numerical solutions. At $h=0.01$, the time step $\tau = 0.001$ and the number of particles $N_{total} = 2 \times 990$ are used; at $h=0.001$, the step is $\tau = 0.0001$ and $N_{total} = 2 \times 9900$. For MK method time step is $\tau = 0.0001$ for all spatial resolution.}
    \label{fig:Dustyshock}
\end{figure}

Fig.~\ref{fig:Dustyshock} shows gas and dust velocity as the solution of Dustyshock problem at the time moment $t = 0.2$. The same methods MK, ISPH, IDIC as in Fig.~\ref{fig:Dustywave} are used. For $h=0.01$ we take $N_{total} = 2 \times 990$, $\mathrm{CFL}=0.1$, for $h=0.001$ --- $N_{total} = 2 \times 9900$, $\mathrm{CFL}=0.1$ 

One can see that the numerical solutions obtained by MK and ISPH methods with the smoothing length increased from $h=0.001$ (the condition (\ref{eq:hLaibePrice}) is satisfied) to $h=0.025$ ((\ref{eq:hLaibePrice}) is violated) acquire a pronounced dissipation. The observed tendency to solution dissipation was described in \cite{BateDust2014,LaibePrice2011}. We can see from Figs.~\ref{fig:Dustywave},~\ref{fig:Dustyshock} and Tables~\ref{tab:L2dustywave},~\ref{tab:L2dustyshock} that the maximum level of dissipation is obtained in the case of explicit schemes MK. As follows from the central panel in Fig.~\ref{fig:Dustywave}, the semi-implicit ISPH scheme gives a smaller dissipation at the same smoothing length as compared to MK. Moreover, thanks to semi-implicit approximation of the drag force, the ISPH scheme has no restrictions on the time step (\ref{eq:tauStiff}). 

MK and ISPH are the fully Lagrangian methods, which means that all forces are calculated without the introduction of a spatial grid. The IDIC method is a combination of Lagrangian and Euler approaches because the drag force is computed using the decomposition of particles into Euler volumes. One can see on the right upper panel of Fig.\ref{fig:Dustywave} that these numerical solutions obtained by IDIC are free of dissipation, and at $h=0.025$ the wave amplitude is reproduced without visible error, in distinction to MK and ISPH. 

\label{sec:results}
\begin{table}
\small
\caption{Error for dust velocity computed with 3 different methods in $L_2$ norm for DUSTYWAVE problem.  $N_{total}=2\times600$ SPH particles.}
\begin{center}
\begin{tabular}{|c|c|c|c|c|}
\hline & & MK & ISPH & IDIC \\
\hline $h=0.01$ & $\tau=0.00025$ & 0.0417 & 0.0332 & 0.0003 \\
\hline $h=0.02$ & $\tau=0.001$ & 0.1486 & 0.079 & 0.0012 \\
\hline
\end{tabular}
\end{center}
\label{tab:L2dustywave}
\end{table}

\begin{table}
\small
\caption{Error for dust velocity computed with 3 different methods in $L_2$ norm for DUSTYSHOCK problem. $N_{total}=2\times990$ SPH particles.}
\begin{center}
\begin{tabular}{|c|c|c|c|c|}
\hline & & MK & ISPH & IDIC \\
\hline $h=0.01$ & $\tau=0.000025$ & 0.1177 & 0.2101 & 0.0457 \\
\hline $h=0.02$ & $\tau=0.0001$ & 0.1786 & 0.29 & 0.0605 \\
\hline
\end{tabular}
\end{center}
\label{tab:L2dustyshock}
\end{table}

\section{SUMMARY}
\label{sec:summary}
Simulation of the dynamics of gas-aerosol particle mixtures is computationally challenging, especially in particle methods as Smoothed particle hydrodynamics. In the paper we compared ability of fully lagrangian methods MK \cite{MonaghanKocharyan1995} and ISPH \cite{BateDust2014} and euler-lagrangian method IDIC \cite{SPHIDIC} to reproduce asymptotical properties of the solution for dynamics of gas-dust mixtures with intense interphase interaction. We found that IDIC method where drag is computed using euler cells is asymptotic preserving and allows to use timestep and smoothing length independent on drag intensity. MK and ISPH methods require fine spatial resolution for intense interphase interaction.

\textit{Acknowledgements.} This work was supported by the Russian Science Foundation grant 19-71-10026.

\end{document}